%% file: main2.tex
\documentclass{nature2}
\usepackage{graphicx}
\usepackage{amsmath}
\usepackage{amssymb}
\usepackage{color}
\usepackage{verbatim}
\usepackage{tabu}
\usepackage{sansmath}
\usepackage{upgreek}
\usepackage{epsfig}
\usepackage{caption}
\usepackage{url}
\usepackage[dvipsnames]{xcolor}

\DeclareCaptionLabelFormat{adja-page}{\hrulefill\\#1 #2 \emph{(previous page)}}

\def\lapp{\ifmmode\stackrel{<}{_{\sim}}\else$\stackrel{<}{_{\sim}}$\fi}
\def\gapp{\ifmmode\stackrel{>}{_{\sim}}\else$\stackrel{>}{_{\sim}}$\fi}

\def\ltsima{$\; \buildrel < \over \sim \;$}
\def\lsim{\lower.5ex\hbox{\ltsima}}
\def\gtsima{$\; \buildrel > \over \sim \;$}
\def\gsim{\lower.5ex\hbox{\gtsima}}

\newcommand{\be}{\begin{equation}}
\newcommand{\en}{\end{equation}}

\title{Finding high-redshift gamma-ray bursts in combined near-infrared and optical surveys}

\bigskip

\author{
S. Campana$^{1}$, \allowbreak
G. Ghirlanda$^{1,2}$, \allowbreak
R. Salvaterra$^{3}$, \allowbreak
O.A. Gonzalez$^{4}$, \allowbreak
M. Landoni$^{1}$, \allowbreak
G. Pariani$^{1}$, \allowbreak
A. Riva$^{5}$, \allowbreak
M. Riva$^{1}$, \allowbreak
S.J. Smartt$^{6}$, \allowbreak
N.R. Tanvir$^{7}$, \allowbreak
S.D. Vergani$^{8}$ 
}
\newcommand{\affils}{
\begin{affiliations}
\item{INAF - Osservatorio astronomico di Brera, Via E. Bianchi 46, I-23807, Merate (LC), Italia}
\item{INFN – Sezione di Milano-Bicocca, Piazza della Scienza 3, I-20126 Milano, Italia}
\item{INAF - 
Istituto di Astrofisica Spaziale e Fisica cosmica,
via Alfonso Corti 12, I-20133 Milano, Italia}
\item{STFC UK Astronomy Technology Centre, The Royal Observatory Edinburgh, EH9 3HJ, Edinburgh, UK}
\item{INAF - Osservatorio astrofisico di Torino, Strada Osservatorio 20, I-10025, Pino Torinese (TO), Italia}
\item{Astrophysics Research Centre, School of Mathematics and Physics, Queen's University Belfast, Belfast BT7 1NN, UK}
\item{School of Physics and Astronomy, University of Leicester, University Road, Leicester, LE1 7RH, UK}
\item{GEPI, Observatoire de Paris, PSL University, CNRS, 5 Place Jules Janssen, F-92190, Meudon, France}
\end{affiliations}
}
\begin{document}

\maketitle

\affils

The race for the most distant object in the Universe has been played by long-duration gamma-ray bursts (GRBs), star-forming galaxies and quasars.
GRBs took a temporary
lead with the discovery of GRB 090423 at a redshift $z=8.2$$^{1,2}$,
but now the record-holder is the galaxy GN-z11 at $z=11.0$$^{3,}$\footnote{{The recently released James Webb Space Telescope first images have revealed a number of candidates in $z\sim 9-15$ (e.g., [4]).}}.
However, galaxies and quasars are very faint (GN-z11 has a magnitude $H=26$), hampering the study of the physical properties of the primordial Universe.
On the other hand, GRB afterglows are brighter by a factor of $\gsim 100$, but with the drawback of being transients, and lasting only for 1--2 days.
%
When bright (e.g., $H=18.6$ mag at 7 hr from the explosion as observed for GRB 090423$^{1,2}$),
GRB spectroscopic afterglow studies could provide independent constraints on the reionisation history,
on the ionising photon escape fraction
and on the cosmic history of metal enrichment.
GRBs could be used to select high-$z$ galaxies independently from their brightness, thus sampling the galaxy luminosity function below the sensitivity limit of deep surveys.
They also are expected to provide important information about the nature of the stellar population in the first galaxies by, e.g., identifying explosions from the first population of metal-free stars (the so-called PopIII stars), showing the presence of their distinctive metal enrichment signature
and measuring the stellar initial mass function evolution at early epochs.

The major limitation
is that, with current instrumentation, high-redshift ($z\gsim 6$) GRBs are extremely rare, with just 9 events recognised in 
17.5 years of {\it Swift} observations$^{5}$.
This is mainly due to the limited sensitivity and field of view of current (the Neil Gehrels {\it Swift} observatory) and soon to be launched (Space-based multi-band astronomical 
Variable Objects Monitor, {\it SVOM}) space missions, and to the relatively faint nature of high-redshift GRBs. It is also likely that some of these high-redshift GRBs are lost during the, sometimes patchy, follow-up process. After the GRB alert sometimes the afterglow is followed up only at optical wavelengths or the telescopes for the near-infrared (nIR) spectroscopic follow-up are not available or weather conditions do not allow for rapid observations.

Here we describe a novel approach to the discovery of high-redshift ($z\gsim 6$) GRBs, exploiting their nIR emission properties. 
The afterglows of high-redshift GRBs are naturally absorbed, like any other source, at optical wavelengths by hydrogen along the line of sight in the intergalactic medium (Lyman-$\alpha$ absorption).
We propose to take advantage of the deep monitoring of the sky by the Vera Rubin Observatory$^{6}$,
to simultaneously observe exactly the same fields with a dedicated nIR facility.
By comparing the two streams of transients, one can pinpoint transients detected in the nIR band and not in the optical band.
Fast transients detected only in the nIR and with an AB colour index $r-H\gsim 3.5$ are high-redshift GRBs, with a low contamination rate (see Box).

In order to estimate the number of nIR high-redshift GRBs, 
we carried out simulations using a population synthesis code$^{7}$.
A population of long GRBs is simulated based on their luminosity function (i.e., the number of sources as a function of their luminosity or energy) and their formation rate as a function of  
redshift (i.e., cosmic rate density). 
The free parameters of these functions have been derived by reproducing 
the observed properties of well selected samples of GRBs collected in the past 15 years. 
In particular, the population is calibrated by 
reproducing the fluence, peak flux, observer frame peak energy and observer-frame duration distribution of GRBs 
detected by {\it Fermi} and {\it Swift}. In order  to minimise the impact of observational biases, which always play a role in GRB studies, a flux-limited,  $97$\% complete 
in redshift, sub-sample of bright GRBs detected by {\it Swift}$^{8}$
was used. This sample provides constraints on the GRB rate density. 
Long-duration GRBs are associated with the core-collapse of massive stars,
thus indicating they might be  good tracer of the star formation. 
However, theoretical arguments indicate that GRBs require a low-metal content in the progenitor$^{9}$,
resulting in an increase of GRBs with redshift over the nominal star formation rate.
This has been addressed including a GRB density evolution with redshift$^{7,8}$.

In addition to the GRB prompt properties, the population code simulates 
the afterglow emission based on the  model by Ryan et al.$^{10}$
for a fireball decelerating in a constant density external medium (see Table \ref{param} for the adopted parameter values). 
The afterglow luminosity depends on the kinetic energy of the jet and on the shock efficiencies in amplifying the magnetic field and 
accelerating the emitting particles and it has been demonstrated that there is a close relationship between the afterglow luminosity and 
the prompt-emission high-energy luminosity$^{11}$.
This code has been adopted for rate estimates for the Transient High-Energy Sky and Early Universe Surveyor ({\it THESEUS})$^{12}$
and the {\it Gamow Explorer}$^{13}$.

Having set the parameters for a MonteCarlo simulation, we run the code asking what is the rate of GRBs as a function of redshift for a given limiting magnitude
and at a given time after the GRB event. This is shown in Fig. 1 (top panel) for $H<21$ (AB magnitude system). The decay of the GRB rate as a function of redshift for $z>5$ follows a power-law evolution $\propto z^{-3.4}$ (solid line in Fig. 1 - top panel). Owing to the power-law temporal decay of the afterglow light curve, the data points sampled at different epochs 
follow a similar $z$-decay. 
The dependence of the rate of high-redshift ($z\gsim6$) GRBs on the limiting $H$ magnitude is shown in Fig. 1 (bottom panel). 
We estimate that about 1/3 of high-redshift {\it Swift} GRBs are missed during the follow-up process. 
We also note that the population sampled by nIR observations is somewhat different from the one sampled by current high-energy instruments. Fig. \ref{fig:eiso-z} shows the contour levels in plane of the prompt emission isotropic GRB energy $E_{\rm iso}$ versus redshift for the simulated long GRB population detectable with $H<21$. For comparison the nine long GRBs known with redshift $z>6$ are shown (star symbols). The nIR detection samples relatively less energetic events with respect to those currently detected by, e.g., {\it Swift} and in the near future {\it SVOM}. 

The high-$z$ nIR detection rates are indeed very promising. The only problem is identifying these transients as GRB afterglows. 
Lyman-$\alpha$ absorption comes to the rescue, heavily absorbing radiation at optical wavelengths for  high-redshift objects.
If we can simultaneously observe in the optical and in nIR bands, transients detected only at longer wavelengths are natural candidates to be high-$z$ GRBs.
We can make these candidates stronger by increasing the colour index, i.e., the magnitude difference between optical and nIR, 
ensuring the optical observations are sufficiently deep. 

In the next few years, we will have of the most formidable optical survey machine: the Vera Rubin Observatory$^{6}$.
The Rubin Observatory will carry out the Legacy Survey of Space and Time (LSST), reaching a limiting magnitude $r\sim 24.5$ in 30 s exposures (AB magnitude system, single visit, $5\,\sigma$).
Here we propose the construction of a new nIR facility to observe in tandem with the Rubin LSST to unveil the high-redshift Universe
through the identification of GRBs. To be conservative, we consider a telescope and nIR camera able to reach magnitude $H=21$ in 30 s, with the same
field of view as the Rubin telescope (9.6 deg$^2$). This will allow us to look for transients with $r-H\gsim 3.5$, enough to assure that the transients are 
at high redshift$^{14,15,1,2}$.
Then, we assume 10 hr observations per night, $80\%$ of good weather nights, 30 s exposures, and $73\%$ open shutter time as 
a fraction of the observing time$^{16}$.
Based on these metrics, we estimate that we would detect $\sim 11.0$ GRBs per year at $z\gsim 6$. The corresponding rate at $z\gsim 10$ is $\sim 2.7$ GRBs per year. 

A nIR telescope with the required sensitivity and field of view does not currently exist. It should have the same field of view of the Rubin telescope (9.6 deg$^2$) and should reach magnitude $H=21$ in 30 s (in two sub-exposures of $\sim 15$ s). The system should be optimised to observe in the nIR to improve performance and minimise the telescope diameter. In Table \ref{telescope}, we provide a summary of the main existing telescopes with nIR imaging instruments. It is readily apparent that, if the system is not nIR optimised, an 8m-class telescope is needed. Using the Advanced Exposure Time Calculator$^{17}$ 
tool (http://aetc.oapd.inaf.it/), we simulated astronomical images with customisable telescope and instrument combinations. 
Adopting state of the art detectors (efficiency $>0.8$, low readout noise), we end up with a $\sim 5$m-telescope. We can reduce the need for a large mirror, by slightly widening the observing band: extending the short wavelength edge of the $H$ band (1320-1940 nm) to 1220 nm (the central wavelength of the $J$ band), would reduce the aperture requirement to a $\sim 3$m-telescope. A dedicated trade-off study would be needed to optimise the wavelength range. Assuming a 4m telescope with an optical design similar to Rubin's telescope (as a baseline), we can roughly estimate a cost of 35-40 MEuro for the (robotic) telescope and 5-7 MEuro for the dome and the infrastructure, on the market. No particular requirements on the pointing speed are needed, as this telescope will slavishly mirror Rubin's activity. To minimse the camera costs, we can foresee the use of InGaAs detectors with a $\sim 0.8-1$ arcsec/pixel scale. This camera would be the largest nIR camera ever built, even larger than the VISTA IR Camera (VIRCAM)$^{18}$. 
Based on the VIRCAM experience, the estimated cost is 10-13 MEuro (depending on the pixel scale).
No adaptive optics corrections are foreseen. The camera would comprise 128-200 Mpixels, resulting in 2-3 times the VIRCAM data rate. This nIR telescope must be located in the Southern hemisphere, in order to work in tandem with the Rubin telescope. Favourable sites would be ESO's La Silla and Paranal or US's Cerro Tololo and Pachon.

High-redshift GRB afterglows can be directly extracted by comparing the stream of nIR transients with the (public) stream of Rubin LSST transients. Rapid forced photometry on the LSST data 
will ensure the objects selected are optically faint or undetected.
We stress that this method is independent of high-energy facilities, such red and fast objects are unique in astronomical transient streams, and it will allow us to sample a less-extreme population of high-$z$ GRBs (see Fig. 2).
Transients detected in the nIR and missing from the Rubin LSST stream are selected for follow-up observations.
To take advantage of these detections, dedicated nIR spectroscopic facilities (such as X-shooter and SOXS at ESO or FLAMINGOS-2 at the Gemini, in the Southern hemisphere) should confirm the candidates, letting then
the largest next-generation facilities, such as the James Webb Space Telescope ({\it JWST}), the extremely large telescopes (ELT, TMT) at nIR wavelengths, 
the Advanced Telescope for High ENergy Astrophysics ({\it Athena}) at X-ray energies, or the Square
Kilometre Array (SKA) in the radio, to point at them disclosing the line of sight and the host galaxy properties.


\clearpage

\begin{table*}[!htb]
    \centering
    \begin{tabular}{l|ll}
    \hline
            & $\mu$ & $\sigma$ \\
    \hline
$\mathfrak{G}[\log \epsilon_e]$  &  --1.0  &  0.1 \\
$\mathfrak{G}[\log \epsilon_B]$  &  --3.0  &  0.1 \\
$\mathfrak{G}[\log n]$  & 0.0   & 0.5 \\
$\mathfrak{G}[p]$  & 2.1 & 0.2 \\
    \end{tabular}
    \caption{
The afterglow model adopted for simulating the GRB population has a set of free parameters. Among these, the shock micro-physical parameters ${\bf f}=[\epsilon_B,\epsilon_e]$, corresponding to the fraction of energy shared by the shock between particles ($\epsilon_e$) and the magnetic field ($\epsilon_B$). The circum-burst number density is parameterised with $n$.
These parameters were assigned by sampling Log--normal distributions $\mathfrak{G}[\log \bf{f}|\mu,\sigma]$. The index of the energy distribution of shock accelerated electrons was sampled from a Gaussian distribution $\mathfrak{G}[p|\mu,\sigma]$. These are typical afterglow parameter values and it has been verified that they produce nIR fluxes consistent with the ranges observed in the few high-redshift GRBs known. 
To account for the possibility that the very early afterglow emission is characterised by a rising phase, the deceleration time has been computed by assigning to each GRB a bulk Lorentz factor $\Gamma_0$, sampled from the empirical correlation between $\Gamma_0$ and the prompt emission isotropic energy $E_{\rm iso}$ (ref. 19).
We assume negligible dust absorption, as indicated by high-$z$ GRB afterglow observations$^{20}$, 
that in all cases show $A_V<0.15$. 
This is in line with the expected proprieties of high-$z$ GRB host galaxies$^{21}$ 
and the general trend of decreasing dust absorption 
in GRB afterglows with redshift$^{22}$.
}
    \label{param}
\end{table*}

\clearpage

\begin{table*}[!htb]
    \centering
    \begin{tabular}{ll|r}
    \hline
Telescope & Instrument & Total Exposure \\
          &            & time to SNR=5 \\
    \hline
TNG (3.6m)    & NICS      & 190 s\\
UKIRT (3.8m)  & WFCAM     & 132 s\\
CFHT (3.6m)   & WIRCAM    & 114 s\\
NTT (3.6m)    & SOFI      & 84 s \\
GeminiN (8.1m)& NIRI      & 75 s\\
VISTA (4.1m)  & VIRCAM    & 70 s \\
GeminiS (8.1m)& FLAMINGO2 & 60 s\\
LBT (8.4m)    & LUCI      & 47 s\\
Keck (10m)    & OSIRIS    & 40 s\\ 
\hline
Subaru (8.3m) & MOIRCS    & 18 s \\
GTC (10.4m)   & CIRCE     & 8 s \\
VLT (8.2m)    & HAWK-I    & 8 s \\
\hline
    \end{tabular}
\caption{Exposure time needed to reach a Signal to Noise Ratio of 5 in nIR exposures of different telescopes and instruments. Values have been obtained using the relevant exposure time calculator with a 0.5 Full Moon Illumination, 1.2 airmass, 10 mm Precipitable Water Vapour, $1''$ seeing and a flat power law spectrum with index 0. We used different DIT combinations, being 2 exposures lasting 15 s the preferred choice.}
    \label{telescope}
\end{table*}

\clearpage

\begin{figure*}[!th]
\begin{center}
\vskip -4truecm
\includegraphics[width=12cm]{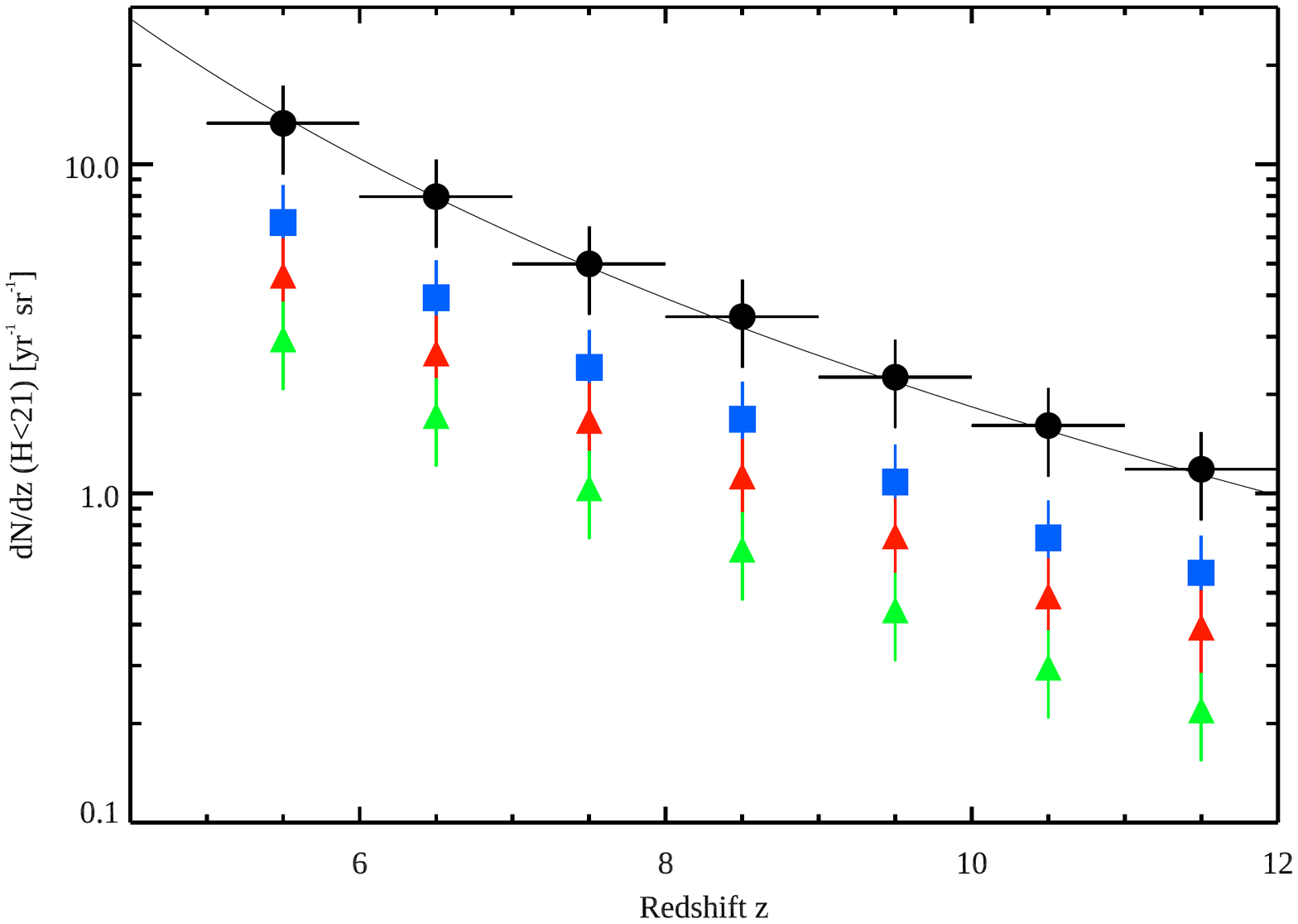}
\vskip -9truecm
\includegraphics[width=13.3cm]{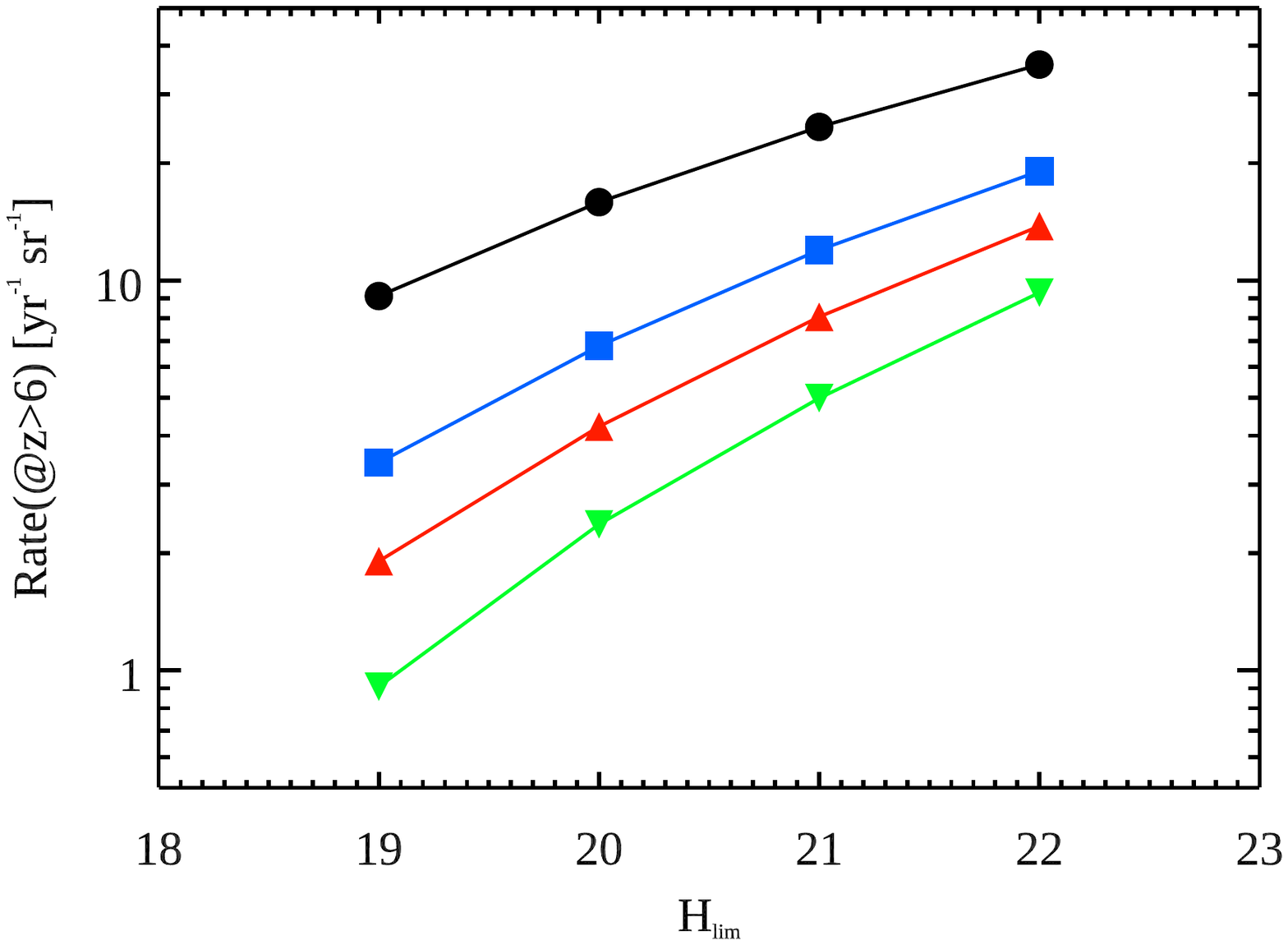}
%
\vskip -4truecm
\caption{{\bf Rates of high-redshift GRBs depending on the nIR limiting magnitude and redshift.}
Top: 
predicted rate per unit solid angle of $H<21$ high-redshift ($z\gsim 6$) GRB afterglows as a function of redshift and at different observing times (black-circle: 1 hr; blue-square: 6 hr; red-triangle: 12 hr; green-triangle: 24 hr). For reference a decay $\propto z^{-3.4}$ is shown by the solid black curve. Horizontal bars (shown only for the circle symbol) represent the redshift bin width. 
Bottom: rates of high-redshift GRBs ($z\gsim 6$) as a function of the limiting $H$ magnitude in the AB system at a fiducial times (same symbols as in the top panel) after the prompt.
}
\label{fig:rates}
\end{center}
\end{figure*}

\begin{figure}
    \centering
    \vskip -3truecm
    \includegraphics[width=15cm]{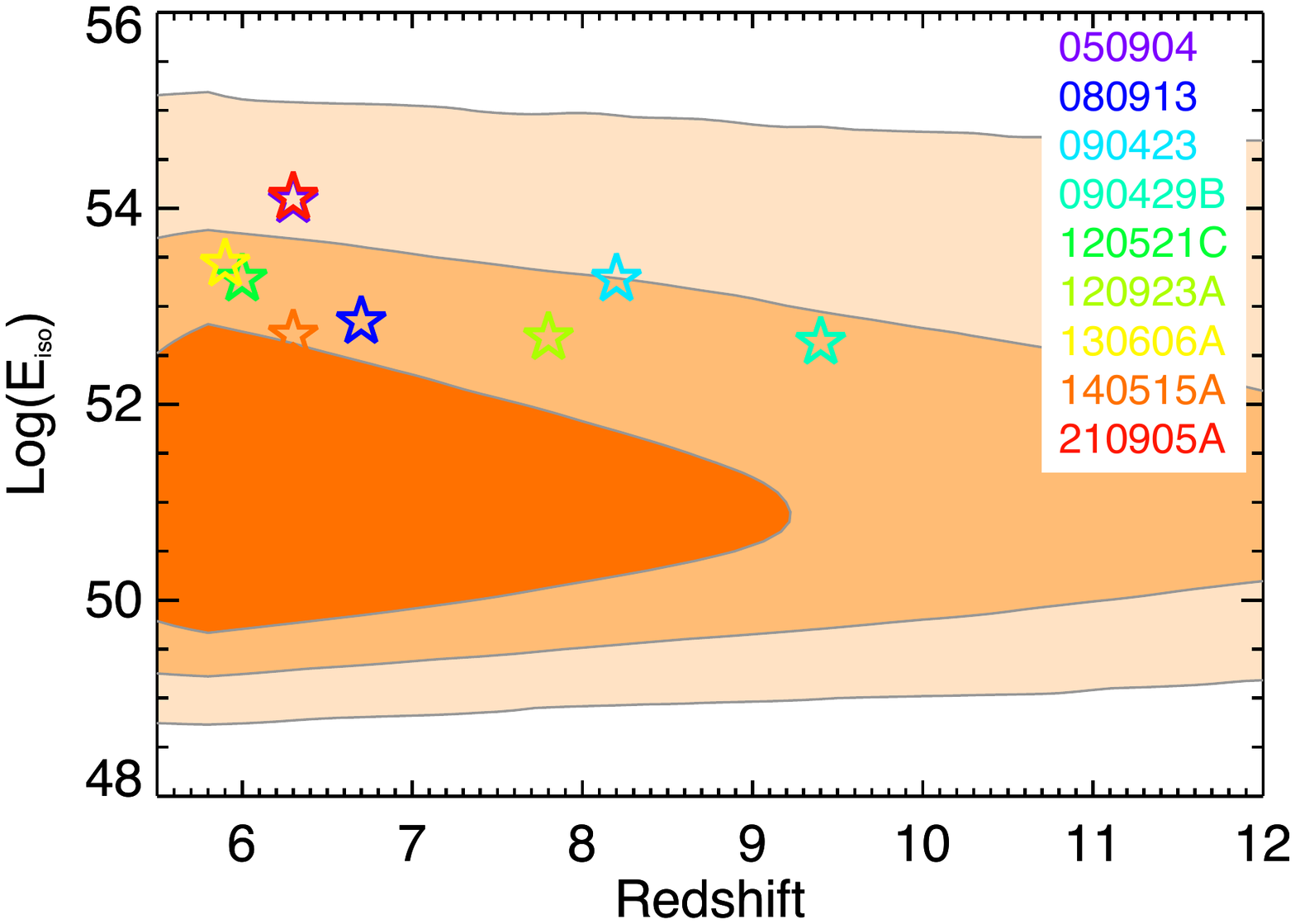}
    \vskip -3truecm
    \caption{Prompt emission isotropic equivalent energy versus redshift of the simulated population of events with $H<21$ (shaded regions showing the 68, 90, 99\% confidence contours). The nine GRBs at $z>6$ are shown with star symbols.}
    \label{fig:eiso-z}
\end{figure}

\clearpage

\begin{addendum}

\item[Acknowledgments]
SC thanks M.G. Bernardini, P. D'Avanzo, A. Melandri, P. Schipani, G. Tagliaferri, and F. Vitali for useful conversations.

%
\item[Competing Interests] The authors declare that they have no competing financial interests.
    
    
    \textbf{Correspondence and requests for
        materials} should be addressed to Sergio Campana\\ (email: sergio.campana@inaf.it).
        
\end{addendum}

\input{biblio.tex}

\clearpage

\subsection{BOX: Other red transients (contaminants).} ``Pure" nIR transients are rare and occur in the most heavily absorbed regions of our Galaxy. These can be easily avoided. In addition, heavily absorbed transients in nearby galaxies (e.g., magnetar flares) can be avoided too, being part of a known object catalogue.

$\bullet$ Some classes of Young stellar objects (YSOs) show powerful nIR outbursts. These systems are low-mass pre-main sequence objects.
FU Orionis-type stars (FUors) experience 4-5 mag outbursts lasting several decades; EX Lupi-type stars (EXors) show dimmer outburst (2–3 mag) which last for a few months-years$^{23}$.
Their quiescence counterpart is usually brighter than $H\sim 21$ and they are embedded in Galactic star-formation regions. Despite the heavy absorption, they are often observed at optical wavelengths, too. We should be able to discard them, based on pre-existing optical and nIR counterparts, as well as on their location in Galactic star-formation regions.

$\bullet$ Late M-L dwarfs have usually very red colours, becoming redder as one progresses along the spectral sequence. Many of these M-L dwarfs show prominent flares, because they retain their magnetic field and remain active for Gyr. Fortunately, flares are significantly hotter than the star's temperature. 
They are detectable in white light, often with a prominent H$\alpha$ line, as testified, e.g., by the optical ASAS-SN survey$^{24}$.
The  flare spectra of ultra cool M-dwarfs have been modelled as black body emission of $\sim 10,000$ K$^{25}$.
The investigation of hundreds of M-dwarf flares from 
the ATLAS survey
shows them to have 
a characteristic exponential decay on a timescale of 
30-60 minutes, with a red stellar counterpart visible in Pan-STARRS, SDSS, UKIDDS, VISTA or WISE archives. Hence, these
Galactic flares are easily distinguished from GRB afterglows. A combination of high Galactic extinction and large flares could potentially result in very red flare colours, but this is easily circumvented, again, by avoiding high-extinction regions. 

$\bullet$ `Dark' GRBs are probably the most severe contaminants for high-redshift GRBs. Naively, dark GRBs are long-duration GRBs with no detectable optical afterglow. A more formal definition involves also the overall strength of the afterglow: a dark GRB is one for which the X-ray to optical spectral slope $\beta_{OX}$ is less than 0.5 [ref. 26]. 
An alternative definition involves also the X-ray spectral slope $\beta_X$, requiring $\beta_{OX}<\beta_X - 0.5$ [ref. 27], 
resulting in a somewhat more restrictive selection (which we adopt here). Dark GRBs are likely caused by intrinsic absorption within the host galaxy$^{28}$,
with a contribution from high-redshift GRBs. These two effects are, however, not immediate to disentangle. 
Using a flux-limited GRB sample complete in redshift Melandri et al.$^{29}$ 
estimated that dark GRBs comprise $\sim 30\%$ of the GRB population. With a different sample, Greiner et al.$^{30}$ 
estimated that $\sim 20\%$ of the dark GRBs are at redshift $z>4-5$. 
In a recent study, Chrimes et al.$^{31}$  
studied a sample of 21 dark GRBs, with {\it Chandra} and {\it HST} data. For $\sim 50\%$ of these dark GRBs, a host galaxy with $r$ magnitude brighter than 25 has been detected. These host galaxies will appear as steady sources in Rubin LSST stacked images, leading to an easy identification. 
\\
Finally, assuming a typical afterglow spectrum with a power law photon index $\Gamma=2$, extending into the optical regime (without a cooling break, resulting in a more restrictive limit), one would expect a $\sim 1$ mag difference from the $r$ and $H$ emission, without absorption. A Galactic extinction curve, which again is a severe approximation being the extinction curve at high-$z$ flatter, provides an absorption of $A_R\sim 0.75$ and $A_H\sim 0.18$, for $A_V=1$.
To gain $\sim 2.5$ from extinction (and satisfy our $r-H\gsim 3.5$ colour selection constraint), one would need $A_V\gsim 4.5$. In the {\it Swift}-BAT6 flux-limited redshift-complete sample$^{8}$, 
we have only 3 GRBs with such a high absorption$^{20,29}$,
indicating that the contribution to extremely dark GRBs able to pass our colour selection criteria (and contaminate high-redshift GRBs), should be at the $\sim 6\%$ level, at the most.
Based on one GRB per day on the whole sky, the number of dark GRBs able to contaminate the search for high-redshift GRBs should be $\sim 4$ GRB/yr, in the worst case, i.e. a factor of two less than the annual rate of high-$z$ GRBs. In passing, we note that dark GRBs are a valuable scientific topic, allowing us to probe star-formation in heavily absorbed environments, too.

\end{document}

%% file: biblio.tex
{\bf References}

\noindent \hangindent2em  \hangafter=1  1.\ Salvaterra, R. et al.,
{\it Nature} {\bf 461}, 1258--1260 (2009).

\noindent \hangindent2em  \hangafter=1  2.\ Tanvir, N.~R. et al.,
{\it Nature} {\bf 461}, 1254--1257 (2009).


\noindent \hangindent2em  \hangafter=1  3.\ Jiang, L. et al.,
{\it Nature Astr.} {\bf 5}, 256--261 (2021).

\noindent \hangindent2em  \hangafter=1  4. Castellano, M. et al., 
{Astrophys. J. Lett.} submitted (arXiv:2207.09436) (2022).

\noindent \hangindent2em  \hangafter=1  5.\ Melandri, A. et al.,
{\it Astron. Astrophys.} {\bf 581}, A86 (2015).

\noindent \hangindent2em  \hangafter=1  6.\ Ivezi{\'c}, {\v{Z}}. et al.,
{\it Astrophys. J.} {\bf 873}, 111 (2019).

\noindent \hangindent2em  \hangafter=1  7.\ Ghirlanda, G. et al.,
{\it Mon. Not. R. Astron. Soc.} {\bf 448}, 2514--2524 (2015).

\noindent \hangindent2em  \hangafter=1  8.\ Salvaterra, R.  et al.,
{\it Astrophys. J.} {\bf 749}, 68 (2012).



\noindent \hangindent2em  \hangafter=1  9.\ Yoon, S.-C., Langer, N. \& Norman, C.,
{\it Astron. Astrophys.} {\bf 460}, 199--208 (2006).


\noindent \hangindent2em  \hangafter=1  10.\ Ryan, G., van Eerten, H., Piro, L. \& Troja, E.,
{\it Astrophys. J.} {\bf 896}, 166 (2020).

\noindent \hangindent2em  \hangafter=1  11.\ Gehrels, N. et al.,
{\it Astrophys. J.} {\bf 689}, 1161--1172 (2008).

\noindent \hangindent2em  \hangafter=1  12.\ Amati, L. et al.,
{\it Exper. Astron.} {\bf 52}, 183--218 (2021).

\noindent \hangindent2em  \hangafter=1  13.\ White, N.~E. et al.,
{\it SPIE} {\bf 11821}, 1182109 (2021).

\noindent \hangindent2em  \hangafter=1  14.\ Cucchiara, N. et al.,
{\it Astrophys. J.} {\bf 736}, 7 (2011).

\noindent \hangindent2em  \hangafter=1  15.\ Tanvir, N.~R. et al.,
{\it Astrophys. J.} {\bf 865}, 107 (2018).

\noindent \hangindent2em  \hangafter=1  16.\ Marshall, P. et al.,
{\it Zenodo} {\bf 10.5281}, 842713 (2017).

\noindent \hangindent2em  \hangafter=1  17.\ Uslenghi, M., Falomo, R., Fantinel, D., 
{\it SPIE}, {\bf 9911}, 99112U (2016).
 
\noindent \hangindent2em  \hangafter=1  18.\ Sutherland, W. et al.,
{\it Astron. Astrophys.} {\bf 575}, A25 (2015).



\noindent \hangindent2em  \hangafter=1  19.\ Ghirlanda, G. et al.,
{\it Astron. Astrophys.} {\bf 609}, A112 (2018).

\noindent \hangindent2em  \hangafter=1  20.\ Covino, S. et al., 
{\it Mon. Not. R. Astron. Soc.} {\bf 432}, 1231--1244 (2013).

\noindent \hangindent2em  \hangafter=1  21.\ Salvaterra, R., 
{\it Journ. High En. Astrophys.} {\bf 7}, 35--43 (2015).

\noindent \hangindent2em  \hangafter=1  22.\ Zafar, T., Watson, D., Fynbo, J.~P.~U., Malesani, D., Jakobsson, P. \& de Ugarte Postigo, A.,
{\it Astron. Astrophys.} {\bf 532}, A143 (2011).

\noindent \hangindent2em  \hangafter=1  23.\ Audard, M. et al.,
in {\it Protostars and Planets VI}, Beuther, H., Klessen, R.~S., Dullemond, C.~P. \& Henning, T., eds. (2014).

\noindent \hangindent2em  \hangafter=1  24.\ Schmidt, S.~J. et al.,
{\it Astrophys. J.} {\bf 876}, 115 (2019).

\noindent \hangindent2em  \hangafter=1  25.\ Kanodia, S. et al.,
{\it Astrophys. J.} {\bf 925}, 155 (2022).



\noindent \hangindent2em  \hangafter=1  26.\  Jakobsson, P., Hjorth, J., Fynbo, J.~P.~U., Watson, D., Pedersen, K.,  Bj{\"o}rnsson, G. \&  Gorosabel, J.,
{\it Astrophys. J.} {\bf 617}, L21--L24 (2004).

\noindent \hangindent2em  \hangafter=1  27.\  van der Horst, A.~J., Kouveliotou, C., Gehrels, N., Rol, E., Wijers, R.~A.~M.~J., Cannizzo, J.~K., Racusin, J. \& Burrows, D.~N.,
{\it Astrophys. J.} {\bf 699}, 1087--1091 (2009).

\noindent \hangindent2em  \hangafter=1  28.\ Campana, S. et al., 
{\it Mon. Not. R. Astron. Soc.} {\bf 421}, 1697--1702 (2012).

\noindent \hangindent2em  \hangafter=1  29.\ Melandri, a. et al., 
{\it Mon. Not. R. Astron. Soc.} {\bf 421}, 1265--1272 (2012).

\noindent \hangindent2em  \hangafter=1  30.\ Greiner, J. et al.,
{\it Astron. Astrophys.} {\bf 526}, A30 (2011).

\noindent \hangindent2em  \hangafter=1  31.\ Chrimes, A.~A., Levan, A.~J., Groot, P.~J., Lyman, J.~D. \& Nelemans, G.,
{\it Mon. Not. R. Astron. Soc.} {\bf 508}, 1929--1946 (2021).